\documentclass[namedreferences]{solarphysics}
\usepackage[optionalrh,solaromanenum]{spr-sola-addons} % For Solar Physics
\usepackage{graphicx}                    % For eps figures, newer & more powerfull
\usepackage{caption}
\usepackage{upgreek}                     % useful mathematical symbols
\usepackage{color}                       % For color text: \color command
\usepackage{lscape}
\usepackage{hyperref}
\begin{document}

\begin{article}
\begin{opening}

\title{Hemispheric asymmetry in the sunspot cycle as a nonextensive phenomenon}
%\shorttitle{Hemispheric asymmetry in the sunspot cycle}

%%%%%%%%%%%%%%%%%%%%%%%%%%%%%%%%%%%%%%%%%%%%%%%%%%%
%% Authors Names
%\author[addressref=aff1,email={e-mail.b@mail.com}]{\inits{F.}\fnm{First~Names}~\lnm{Last~Name~Author-b}}
 \author[addressref={aff1},corref,email={leonardo.pesquisa@gmail.com}]{\inits{L. F. G. }\fnm{Leonardo F. G. }\lnm{Batista}\orcid{0000-0002-2938-180X}}
  \author[addressref={aff2,aff3},email={thiago@fisica.ufc.br}]{\inits{T. M. }\fnm{Thiago M. }\lnm{Santiago}\orcid{0000-0002-6855-8903}}
 \author[addressref={aff2},email={clebersilva@fisica.ufc.br}]{\inits{P. C. F. }\fnm{Paulo C. F. da }\lnm{Silva Filho}\orcid{https://orcid.org/0000-0003-4990-0977}}
 \author[addressref={aff2},email={cleo@fisica.ufc.br }]{\inits{C. V. }\fnm{Cleo V. }\lnm{Silva}\orcid{}}
 \author[addressref={aff2},email={danielbrito@fisica.ufc.br}]{\inits{ D. B. }\fnm{Daniel B. }\lnm{de Freitas}\orcid{0000-0002-8814-6383}}

%%%%%%%%%%%%%%%%%%%%%%%%%%%%%%%%%%%%%%%%%%%%%%%%%%%
%% Runningheads
%
%\runningauthor{}
%\runningtitle{}

%%%%%%%%%%%%%%%%%%%%%%%%%%%%%%%%%%%%%%%%%%%%%%%%%%%
%% Affilations
%% id shold be the same with \author addressref value.

  % \institute{$^{1}$ First affiliation
  %                   email: \url{e.mail-a} email: \url{e.mail-b}\\ 
  %            $^{2}$ Second affiliation
  %                   email: \url{e.mail-c}\\}

\address[id=aff1]{Institute of Aeronautical Technology - ITA, 12228-900, São José dos Campos, SP, Brazil}
\address[id=aff2]{Departamento de Física, Universidade Federal do Ceará, Caixa Postal 6030, Campus do Pici, 60455-900 Fortaleza, Ceará, Brazil}
\address[id=aff3]{Centro de Ciências e Tecnologia, Universidade Federal do Cariri, Av. Tenente Raimundo Rocha, Nº 1639, 63048-080, Juazeiro do Norte, Ceará, Brazil.}

%%%%%%%%%%%%%%%%%%%%%%%%%%%%%%%%%%%%%%%%%%%%%%%%%%%
%%% Abstract
\begin{abstract}
The appearance of dark sunspots over the solar photosphere is not considered to be symmetric between the northern and southern hemispheres. Among the different conclusions obtained by several authors, we can point out that the North-South asymmetry is a real and systematic phenomenon and is not due to random variability. In the present work, we selected the sunspot area data of a sample of 13 solar cycles divided by hemisphere extracted from the Marshall Space Flight Centre (MSFC) database to investigate the behavior of probability distributions using an out-of-equilibrium statistical model a.k.a non-extensive statistical mechanics. Based on this statistical framework, we obtained that the non-extensive entropic parameter $q$ has a semi-sinusoidal variation with a period of $\sim$22 year (Hale cycle). Among the most important results, we can highlight that the asymmetry index $q(A)$ revealed the dominance of the northern hemisphere against the southern one. Thus, we concluded that the parameter $q(A)$ can be considered an effective measure for diagnosing long-term variations of the solar dynamo. Finally, our study opens a new approach to investigating solar variability from the nonextensive perspective.
\end{abstract}

%%%%%%%%%%%%%%%%%%%%%%%%%%%%%%%%%%%%%%%%%%%%%%%%%%%
%% Keywords
%

\end{opening}
%-------------------------------------------------

%%%%%%%%%%%%%%%%%%%%%%%%%%%%%%%%%%%%%%%%%%%%%%%%%%%
%% Sections
%
\section{Introduction}\label{intro}
Among the various solar phenomena studied by humanity, the observation of sunspots is undoubtedly one of the oldest in the history of astronomy. The first sunspot observation data dates back more than 2000 years in China \citep{clark,wittmann}. However, it was only at the beginning of the 17th century that the first observations were made using a telescope, in which Galileo Galilei demonstrated that it was not a planetary transit. Later, \cite{schwabe} deduced the spottedness of the Sun waxed and waned over an 11-year cyclic period. More recently, several studies \citep[e.g.,][]{norton,deng,elborie2016,bada2017,elborie2021} showed that various solar magnetic activity indicators, such as sunspots, faculae, and flares, typically are distributed unevenly between the northern and southern hemisphere of the Sun a.k.a solar hemispheric asymmetry. 

In this context, our work aims to investigate the physical origin of solar cyclic activity to account for the varying configuration of the sun’s magnetic field since it has not yet been understood completely \citep{Hiremath}. In particular, we will explore the dynamics of North-South (hereafter N-S) hemispheric asymmetry as a process predominantly originated from nonlinearity in the solar dynamo \citep{das}.

\subsection{The problem of solar hemispheric asymmetry}
Sunspots are the windows into the Sun's complex interior phenomena which emerge on its surface. They rise as temporary dark spots because they are cooler than their surrounding areas \citep{fd}. In fact, this behavior is due to the strong magnetic activity associated with the differential rotation, in which the plasma flow that accompanies the field lines is compressed to the point of breaking. Mass and energy are then released in a chaotic way, thus generating regions with temperatures around 3700 K. This explains the darkening of these regions on the Sun's surface, comparing the average temperature of 5700 K in the solar photosphere \citep{li}.

Amateur astronomer Richard Carrington revealed an important aspect of sunspots. As reported by \cite{hudson}, Carrington detected from observations made between 1853 and 1861 that the sunspots at different latitudes do not move as a rigid body, but differently as a fluid. This observation paved the way to understand the behavior of spot distributions at different latitudes \citep{hudson}. In this same scenario, \cite{bell} found an N-S asymmetry in the solar area data during Cycles 8 to 18. Generally speaking, N-S asymmetry is a present phenomenon and alternates according to the level of magnetic activity in each hemisphere as described by Cowling's Theorem \citep{cowling}. This behavior has been investigated in terms of several indicators of solar activity, such as the number and area of sunspots, group of sunspots, flares index data, differential rotation, filaments, coronal mass ejection, and photospheric magnetic flux \citep{carbonell}.

In line with this reasoning, \cite{newton1955} studied the N-S asymmetry with the annual values of the sunspot areas from 1874 to 1954. They concluded that the fluctuation in the asymmetry values is not a statistical artifact and that the relative behavior of the two hemispheres is maintained in successive cycles without any indication of change related to the 22-year cycles that refer to the magnetic polarity of the spots. \cite{waldmeier1957,waldmeier1971} investigated the N-S asymmetry between 1955-1969. His works demonstrated that asymmetry is strengthened by a phase difference between the hemispheres that is associated with the eleven-year magnetic cycle. Already \cite{roy} studied N-S asymmetry and areas of large sunspots in the period 1955-1974 and concluded that asymmetry is more evident in patches with a complex magnetic configuration. In addition, \cite{vizoso1989} found that the asymmetry for the northern hemisphere is less evident for extensive spots than that observed in large solar flares, just as the degree of asymmetry does not depend on the area of the spot groups for between 1874 and 1976.

In general, the N-S asymmetry of solar activity has seen mainly investigated by the absolute or normalized difference between the two hemispheres using values of different properties, such as mean sunspot numbers and areas \citep{zou,came}. Several statistical methods and techniques have been applied to extract information related to the complex behavior of the solar activity cycle, such as multifractal analysis, visibility graphs, and wavelet transform, calling for replacing canonical linear statistical approaches with methods originated in the field of nonlinear dynamics \citep[cf.][]{carbonell,Donner,zou,ravi,xu}. However, these methods perform best when the time series obeys a Gaussian-type probability distribution. In this sense, they do not allow an explicit study of long-range correlations found in fat-tailed distributions. A very interesting framework that explains the physical origin of fat tails of probability distributions is grounded on a generalization of Boltzmann-Gibbs entropy also known as Tsallis' nonextensive statistical mechanics \citep{tsallis1988,defreitas2013}. In this context, we chose the nonextensive approach to study the long-term changes in the asymmetry of dynamical characteristics observed at both solar hemispheres' time series \citep{java}.

In the present paper, we study the behavior of the distribution of sunspot areas from cycles 12 to 24 using the non-extensive statistical mechanics to investigate the N-S asymmetry. As highlighted above, the advantages of this statistical approach can be observed by its robustness to quantify the tail of probability distribution, where the effects of gaussianity deviations occur. These effects are related to rare events that emerge the solar magnetic activity and whose behavior controls the dynamics of sunspots. Thus, the aim of our work is to make a detailed study of N-S asymmetry based on the premise that hemispheric asymmetry is a non-extensive phenomenon and that the entropic index $q$ can be a promising parameter to understand the behavior of this asymmetry.

\subsection{Structure of the paper}
This paper is structured as follows. A detailed description of the nonextensive framework and its properties are shown in Section 2. In Section 3, we describe the sample and how we selected 13 solar cycles, as well as their physical implications for the present analyses. Section 4 brings the main results and discussions. Finally, concluding remarks are presented in the last section.

%% Figures
%
\begin{figure}[htb]
\centerline{\includegraphics[width=0.99\textwidth,trim={2.5cm 1cm 2.5cm 2.5cm}]{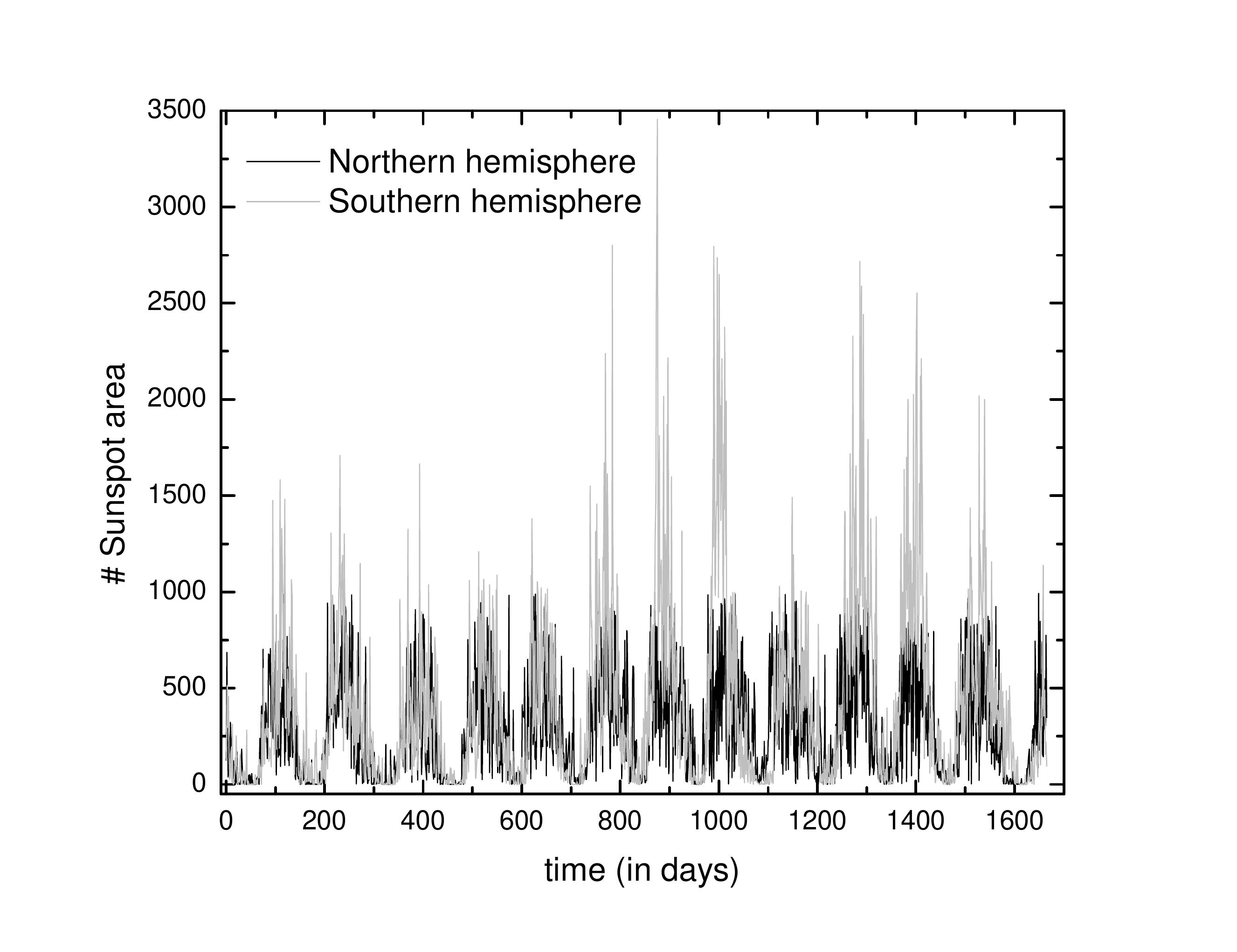}}
\caption{Daily variation of sunspot areas in the Northern and Southern Hemispheres as a function of time for 13 cycles studied from minimum to minimum solar activity. Sunspot areas are given in millionths of the solar hemisphere. The measurements analyzed begin in December 1878 (cycle 12) until December 2019 (cycle 24) and are shown in Table \ref{table2}.}
\label{fig0}
\end{figure}

\begin{figure}[htb]
\centerline{
\includegraphics [width=0.99\textwidth,trim={2.5cm 1cm 2.5cm 2.5cm}]{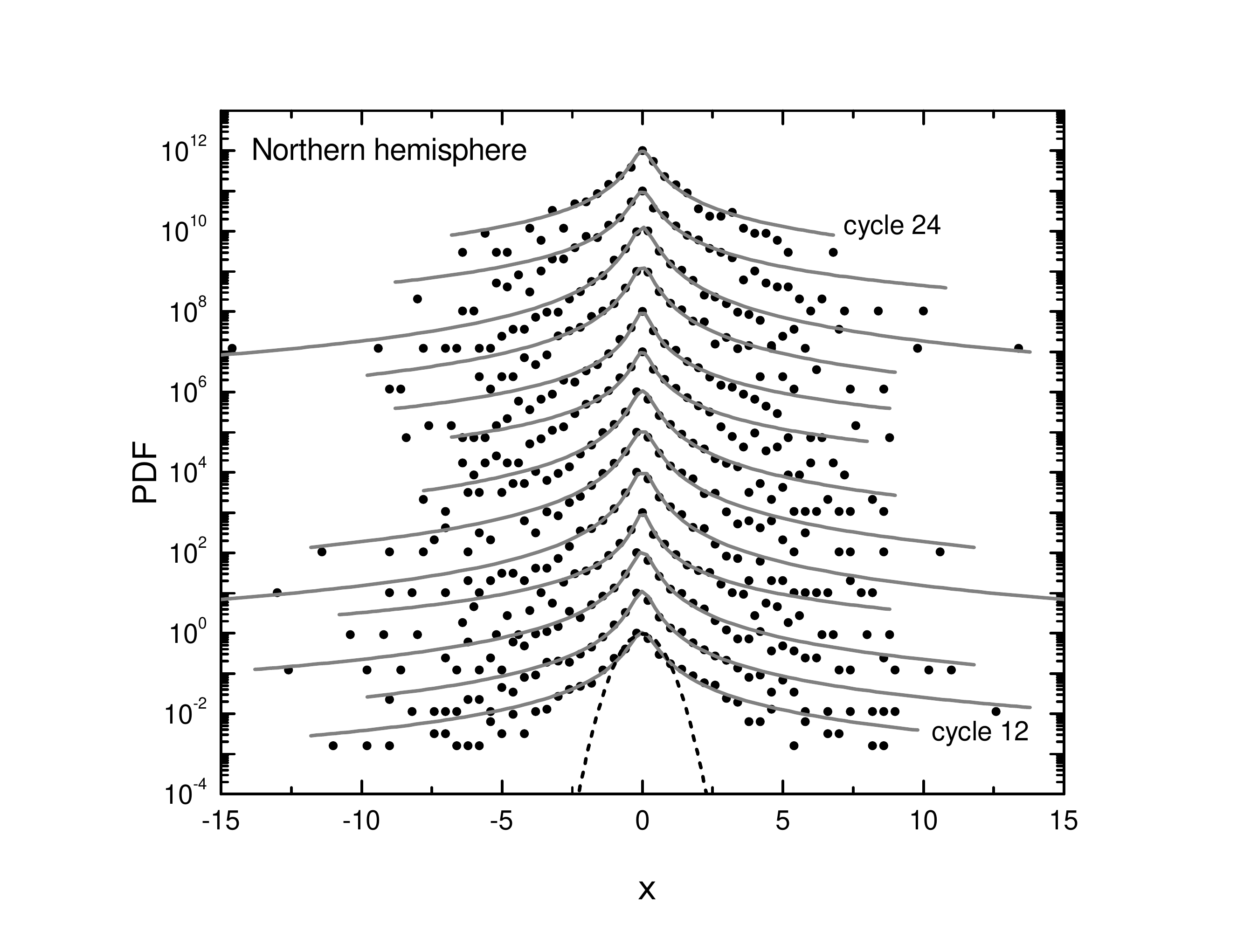}}
\caption{Semi-log plot of the observed (dots) and $q$-Gaussian (full line) PDF of daily sunspot area for the Northern hemisphere from cycle 12 to 24 (from bottom to top, respectively). For the sake of clarity, the distribution functions are shifted up by a factor of 10 each. The standard Gaussian distribution is illustrated only for Cycle 12 (short dashed line).}\label{fig1}
\end{figure}

\begin{figure}[htb]
\centering
\includegraphics [width=0.99\textwidth,trim={2.5cm 1cm 2.5cm 2.5cm}]{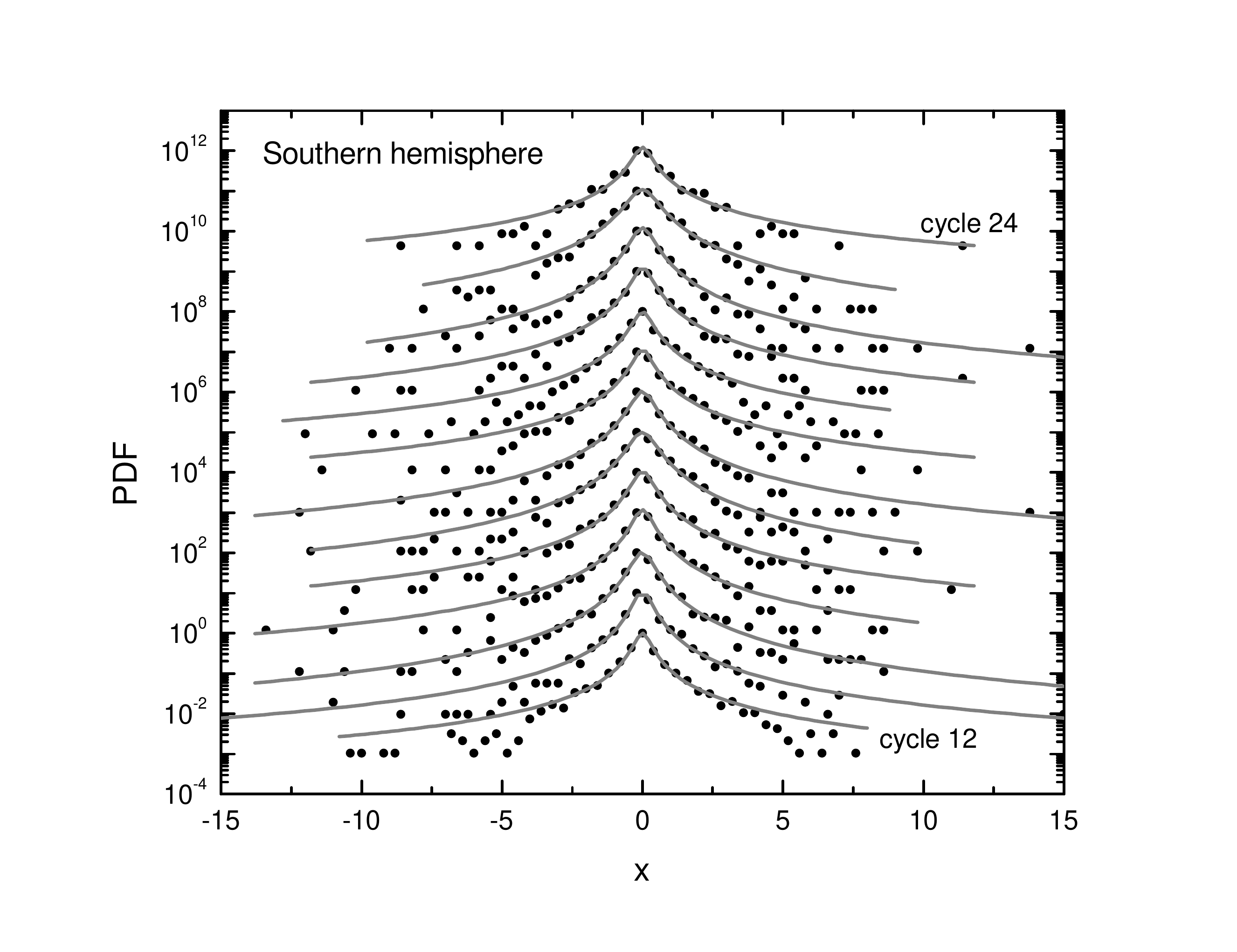}
\caption{Idem Figure~\protect\ref{fig1} for Southern hemisphere. The standard Gaussian distribution not shown.}
\label{fig2}
\end{figure}

\begin{figure}[htb]
\centering
\includegraphics [width=0.95\textwidth,trim={2.5cm 1cm 2.5cm 2.5cm}]{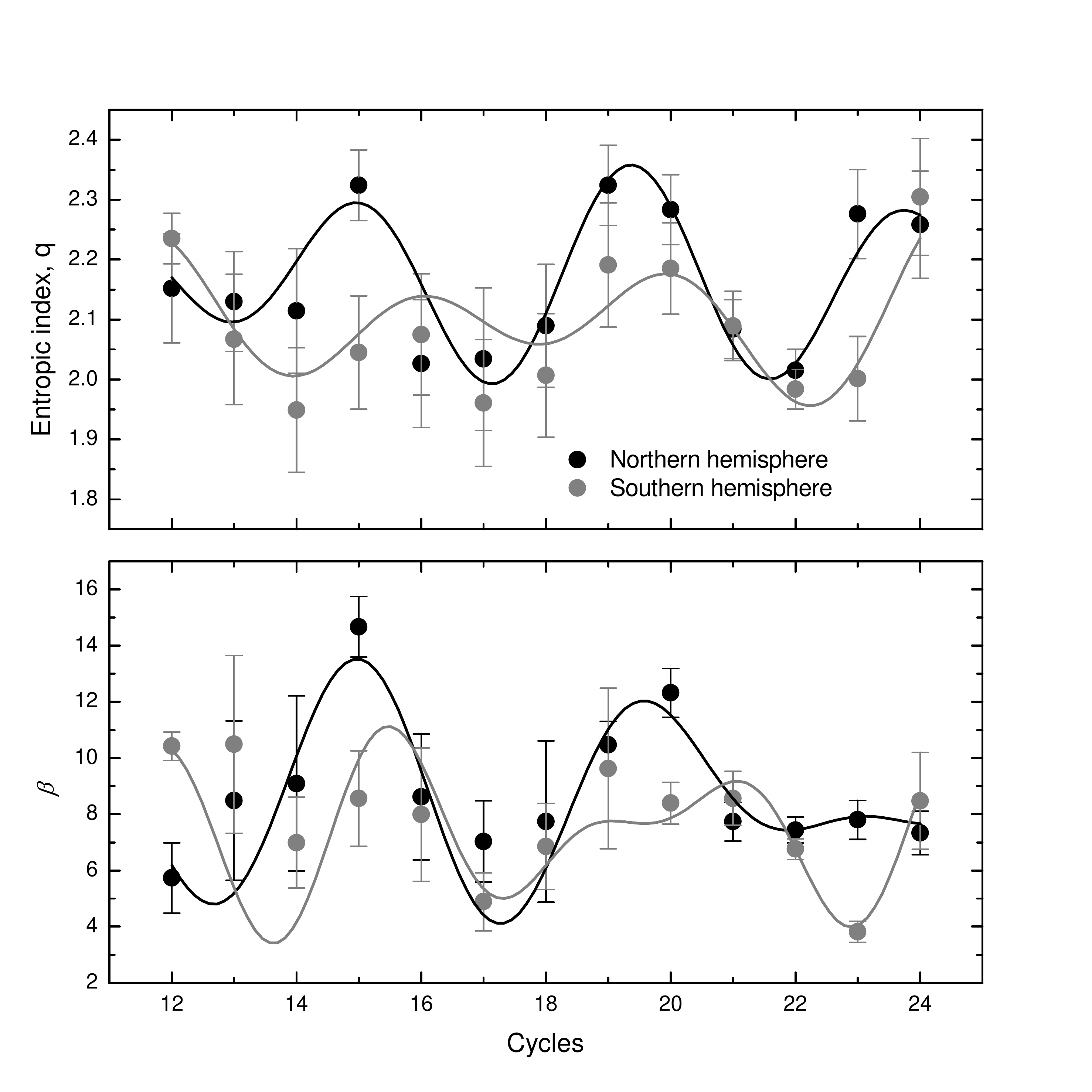}
\caption{Parameters $q$ (top panel) and $\beta$ (bottom panel) are extracted from the Gaussian distribution of sunspot areas as a function of the cycle. The error bars were obtained using the L-M method. There is an evident oscillation with the phase difference between the hemispheres based on $q$- and $\beta$-parameters obtained from the semi-sinusoidal fit. In both cases, the wave period is roughly two cycles.}\label{fig3}
\end{figure}

\begin{figure}[htb]
\centering
\includegraphics [width=0.99\textwidth,trim={2.5cm 1cm 2.5cm 2.5cm}] {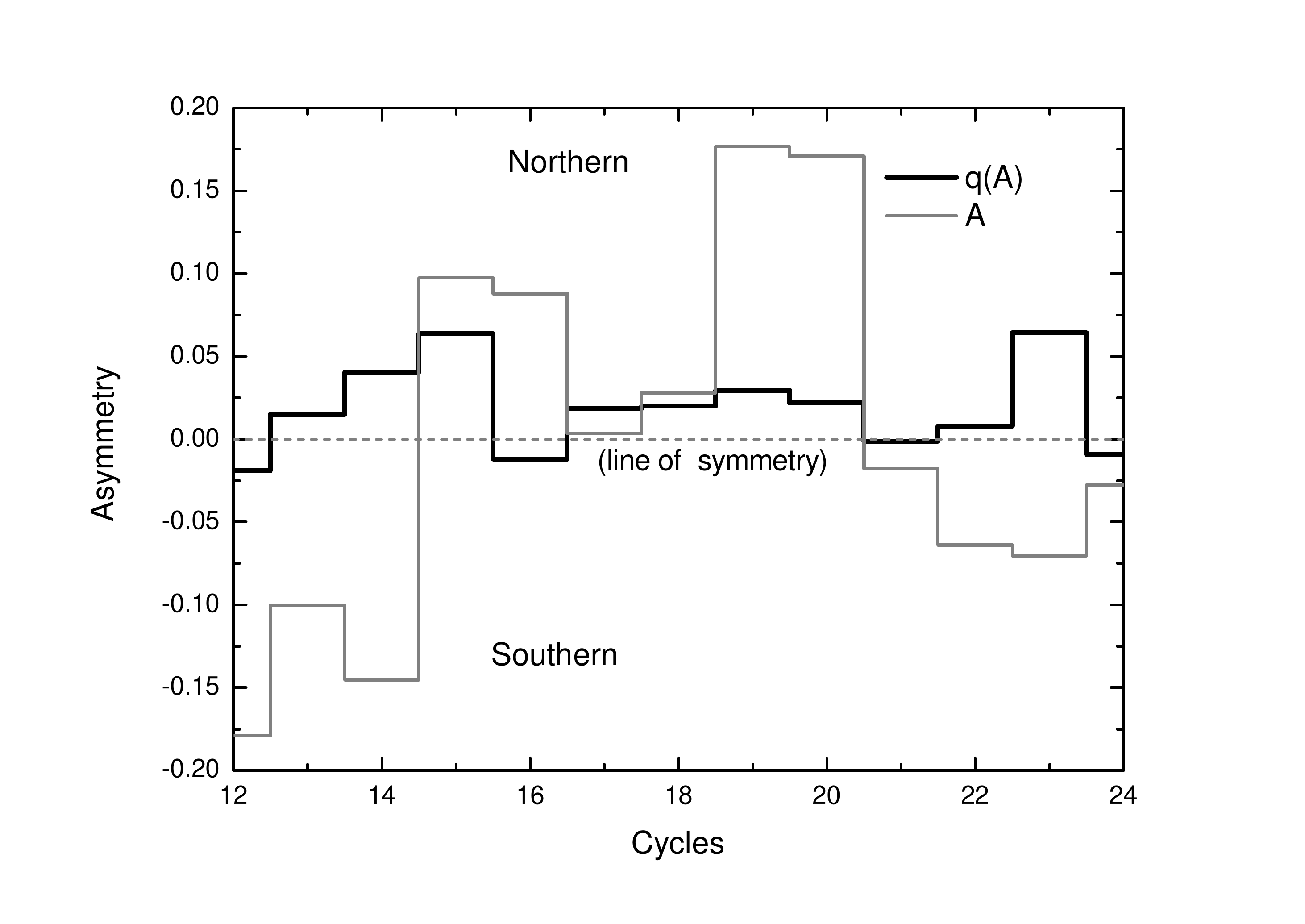}
\caption{The normalized N-S asymmetry $A$ and $q(A)$ of sunspot areas per hemisphere from cycles 12 to 24. The dashed line denotes the line of symmetry when the solar activity of the northern and southern hemispheres is equivalent. Cycles 17 and 21 are the most symmetrical considering both $A$ and $q(A)$. Above the line of symmetry is the region where the northern hemisphere is dominant. The domain below this line denotes the dominance of the southern hemisphere.}\label{fig4}
\end{figure}

\section{Nonextensive formalism}\label{sec:formalism}
An interesting way to investigate the problem of hemispheric asymmetry is to understand how the distribution of the solar indicator, such as sunspot number and sunspot area, behave as a function of the magnetic cycle of $\sim$11 years. In literature, the distributions used for explaining the complex behavior of this problem are based on a usual (boltzmannian) exponential \citep[cf.][]{defreitasetal15,defreitas2020,defreitas2021}. Characterizing the complex behavior of the sunspot dynamics, specifically, in the case of the distribution of the sunspot area, is fundamental to perceive the physical origin of the tail of distributions. Such distributions have an entropy at their core. Among several types of out-of-equilibrium Boltzmann-Gibbs (B-G) entropy that can be found in literature, we can mention the entropies of \cite{druy1930,druy1934}, \cite{ren}, \cite{sharma}, \cite{abe1997}, \cite{papa}, \cite{borges1}, \cite{lands}, \cite{anteneodo1999}, \cite{frank}, \cite{kani} and, finally, the Tsallis non-extensive entropy.

In Tsallis' $q$-entropy the classical statistical formulation is not followed, and entropy additivity does not occur. The new entropic approach is given by $S(A+B)=S(A)+S(B)+(1-q)S(A)S(B)$, where the last term exhibits the interaction between systems $A$ and $B$ that does not exist in the extensive formalism \citep{tsallis1988,Tsallis3,Tsallis4}. In the nonextensive context, entropy is redefined as
\begin{equation}\label{g4}
        S_{q}=k\left(\frac{1-\sum^{W}_{n=1}p^{q}_{n}}{q-1}\right) \quad (q \in\Re),
\end{equation}
where $q$ is the entropic index that characterizes generalization, $k$ is Boltzmann's constant, and $W$ represents the number of microstates in the system. The extra term denotes the interaction between systems A and B.For $q\rightarrow1$ and $p_{n}=1/W$, B-G entropy, defined as $S_{B-G}=k\ln W$, is recovered.

The Tsallis generalized entropy maximization process is responsible for generating the distributions defined as $q$-Gaussians. For this purpose, entropy $S_{q}$ is written in a continuous form as
    \begin{equation}\label{g7}
        S_{q}[p(\mathrm x)]=k\frac{1-\int^{\infty}_{-\infty}\frac{\mathrm d\mathrm x}{\sigma}[\sigma p(\mathrm x)]^{q}}{q-1},
    \end{equation}
where $\mathrm x$ is the dependent variable and $\sigma$ denotes the characteristic length of the problem in question.

For optimizing entropy $S_{q}$, i.e., maximize if $q>1$ and minimize if $q<1$, the following constraints are used
        \begin{equation}\label{g10}
        \int^{\infty}_{-\infty}p(\mathrm x)\mathrm d\mathrm x=1,
    \end{equation}
that corresponds to normalization, as well as
    \begin{equation}\label{g11}
        \left\langle \left\langle \mathrm x^{2}\right\rangle\right\rangle_{q}\equiv\frac{\int^{\infty}_{-\infty}\mathrm x^{2}[p(\mathrm x)]^{q}\mathrm d\mathrm x}{\int^{\infty}_{-\infty}[p(\mathrm x)]^{q}\mathrm d\mathrm x}=\sigma^{2},
    \end{equation}
which corresponds to the variance of $\mathrm x$. In this way, we obtain, starting from the variational problem using the entropic form defined by eq. \ref{g7} plus the links indicated above, Probability Distribution Functions (PDFs) given by
    \begin{equation}\label{g12}
        p_{q}(\mathrm x)=A_{q}[1+(q-1)\beta_{q}\mathrm x^{2}]^{1/(1-q)}, \quad q<3,
    \end{equation}
where the normalization constant $A_{q}$ is obtained for two distinct intervals of $q$,
    \begin{equation}
        A_{q}=\frac{\Gamma\left(\frac{5-3q}{2-2q}\right)}{\Gamma\left(\frac{2-q}{1-q}\right)}\sqrt{\left(\frac{1-q}{\pi}\right)\beta_{q}}, \quad q<1,
    \end{equation}
and,
    \begin{equation}\label{g14}
        A_{q}=\frac{\Gamma\left(\frac{1}{q-1}\right)}{\Gamma\left(\frac{3-q}{2q-2}\right)}\sqrt{\left(\frac{q-1}{\pi}\right)\beta_{q}}, \quad q>1,
    \end{equation}
    where
    \begin{equation}\label{g15}
        \beta_{q}=[(3-q)\sigma^{2}_{q}]^{-1}
    \end{equation}
    and
    \begin{equation}\label{g16}
        \sigma^{2}_{q}=\sigma^{2}\left(\frac{5-3q}{3-q}\right),
    \end{equation}
where $\sigma_{q}$ means the generalized standard deviation as a function of $q$, whereas $\sigma$ is the canonical one.

Before proceeding with entropy optimization, let's take note of the terminology \textbf{impulse}. Impulse is defined as the difference between two neighboring points a time $\tau$ apart. Considering a time series that represents the variability of any parameter $\mathrm X$, the impulse $\rm x$ is given by
    \begin{equation}\label{g8}
        \mathrm x=\mathrm X(t+\tau)-\mathrm X(t).
    \end{equation}
If the interest in question are the fluctuations due to the nearest neighbors, i.e., the level of multifractal noise (where we could find a greater randomness of the studied system), we choose $\tau=1$ \citep{defreitas2021}. The $\tau$-timescale has a unit corresponding to the cadence of the data, that is, if the time series is obtained with data measured day by day, the unit of this parameter is day. The advantage of this procedure is to have the distribution centered at zero. With this, we can also see whether or not there is symmetry between the ``hemispheres'' conjuring up the distributions.

We understand that the behavior of the $q$-index as a function of $\tau$ is very relevant to study the small and large scale fluctuations present in the time series. However, this analysis is outside the scope of our work, since our main interest is to investigate dynamic differences between the solar hemispheres to the noise level. This issue will be addressed in a future communication.

%% Table
   \begin{landscape}

    \begin{table}
     \caption{\footnotesize{List of hemispheric parameters by solar cycles analysed in our work. The cycle number, Start Date, End Date, Duration, and total daily North-South sunspot areas were extracted from MSFC database, whereas the values of $q$ and $\beta$ by hemisphere, shown in the last four columns, were obtained by nonextensive analysis.}}\label{table2}
      \begin{tabular}{ c c c c c c c c c c}
        \\
        \hline		Cycle & Start Date & End Date & Duration & $N$ & $S$  & $q_{N}\pm\delta q_{N}$ & $q_{S}\pm\delta q_{S}$ & $\beta_{N}\pm\delta \beta_{N}$ & $\beta_{S}\pm\delta \beta_{S}$\\
        &  &  & (years) &  &  &  &\\
        \hline
        %\multicolumn{2}{c}{<>}
        12	&	Dez	1878	&	Mar	1890	&	11.3	&	212.2	&	304.7		&	2.15$\pm$0.09		&	2.24$\pm$0.04	&	5.74$\pm$1.25	&	10.42$\pm$0.51	\\
        13	&	Mar	1890	&	Fev	1902	&	11.9	&	271.4	&	331.8		&	2.13$\pm$0.08		&	2.07$\pm$0.11	&	8.48$\pm$2.84	&	10.49$\pm$3.16	\\
        14	&	Fev	1902	&	Ago	1913	&	11.5	&	253.8	&	340.1		&	2.11$\pm$0.10		&	1.95$\pm$0.10	&	9.09$\pm$3.12	&	6.99$\pm$1.61	 \\
        15	&	Ago	1913	&	Ago	1923	&	10	&	361.6	&	297.4		&	2.32$\pm$0.06		&	2.05$\pm$0.09	&	14.67$\pm$1.08	&	8.56$\pm$1.70	\\
        16	&	Ago	1923	&	Set	1933	&	10.1	&	384.4	&	322.2	&	2.03$\pm$0.11	&	2.08$\pm$0.10	&	8.62$\pm$2.23	&	7.99$\pm$2.38	\\
        17	&	Set	1933	&	Fev	1944	&	10.4	&	480.4	&	477		&	2.03$\pm$0.12		&	1.96$\pm$0.11	&	7.03$\pm$1.44	&	4.89$\pm$1.04	\\
        18	&	Fev	1944	&	Abr	1954	&	10.2	&	608.4	&	575.4		&	2.09$\pm$0.10		&	2.01$\pm$0.10	&	7.74$\pm$2.87	&	6.85$\pm$1.53	\\
        19	&	Abr	1954	&	Out	1964	&	10.5	&	838.3	&	586.7		&	2.32$\pm$0.07		&	2.19$\pm$0.10	&	10.47$\pm$0.84	&	9.62$\pm$2.86	\\
        20	&	Out	1964	&	Mar	1976	&	11.7	&	495.4	&	350.8	&	2.28$\pm$0.06	&	2.19$\pm$0.08	&	12.32$\pm$0.86	&	8.39$\pm$0.74	\\
        21	&	Mar	1976	&	Set	1986	&	10.3	&	607.8	&	629.9	&	2.08$\pm$0.05	&	2.09$\pm$0.06	&	7.74$\pm$0.69	&	8.57$\pm$0.96	\\
        22	&	Set	1986	&	Aug	1996	&	9.7	&	549.5	&	624.4	&	2.02$\pm$0.04	&	1.98$\pm$0.03	&	7.44$\pm$0.46	&	6.76$\pm$0.37	\\
        23	&	Aug	1996	&	Dec	2008	&	12.6	&	372.5	&	428.9	&	2.28$\pm$0.07	&	2.00$\pm$0.07	&	7.80$\pm$0.70	&	3.82$\pm$0.37	\\
        24	&	Dec	2008	&	May	2020	&	11.5	&	297.0	&	313.9	&	2.26$\pm$0.09	&	2.30$\pm$0.10	&	7.34$\pm$0.77	&	8.48$\pm$1.72	\\
        \hline
      \end{tabular}
    \end{table}
    \end{landscape}
    
\section{Working sample and asymmetry parameters}
Daily sunspot data from the entire solar disk for each northern and southern hemisphere (NS) were extracted from NASA's Marshall Space Flight Center (MSFC) and compiled by the Royal Greenwich Observatory (RGO) from 1874 to 2016\footnote{The catalog can be downloaded via from the website: \texttt {https://solarscience.msfc.nasa.gov/greenwch/daily$\_$area.txt}}. The data used refers to Cycles 12 to 24. We consider from Cycle 12 onwards simply because the previous cycles had ``gaps'' and ``covers'' in the data, thus avoiding any kind of data interpolation process to homogenize with the other cycles \citep{mursula}. Figure \ref{fig0} shows the time series of the sunspot areas for each cycle separated by hemisphere. In Table \ref{table2}, we insert the main characteristics of this data as the beginning, finish, and duration of the cycle, and total daily sunspot areas by hemisphere.

For the performance of the statistical analysis proposed in the present work, we used the canonical definition of asymmetry $(A)$ as the following relationship \citep{das}:
\begin{equation}
\label{ass1}
A=\frac{N-S}{N+S},
\end{equation}
where $N$ and $S$ respectively are averages of solar magnetic indicators (e.g., sunspot areas, rotation rate, etc.) in the northern and southern hemispheres per cycle. 

In the present paper, we incorporate a new and unique parameter named $q$-Asymmetry Index (hereafter $q(A)$) to mathematically measure the N-S hemispheric asymmetry in sunspots areas defined as:
\begin{equation}
\label{ass2}
q(A)=\frac{q_{N}-q_{S}}{q_{N}+q_{S}},
\end{equation}
where $q_{N}$ represents the nonextensive index of northern hemisphere and $q_{S}$ as that of southern hemisphere. In this way, $q(A)$-index is a measure of the extension of the tail of the distribution, that is, a measure of the deviation from gaussinity. Furthermore, the values of $q(A)$ are calculated for each cycle and hemisphere. With this, as we will see in the next section, we will obtain the profile of $q(A)$ as a function of the solar cycle.

\section{Results and discussions}
We award in Figures \ref{fig1} to \ref{fig2} the distributions of the observational data confronting the fit model through the $q$-Gaussians described by the set of equations \ref{g12} to \ref{g16}. The results of $q$-index and the parameter $\beta$ as a function of solar cycle are cited in table \ref{table2}. These indexes have seen derived from the empirical distribution functions of daily sunspot areas (black circles) as highlighted in Figs. \ref{fig1} and \ref{fig2} for northern and southern hemispheres, respectively. The values of $q$ and $\beta$, and their errors at 0.05 confidence limit, were obtained by a non-linear least-squares minimization using the Levenberg-Marquardt (L-M) algorithm \citep{l,m}, and using the $q$-Gaussian with symmetric Tsallis distribution from eq. \ref{g12}. The measured values of the parameter $q$ strongly suggest that the distribution of sunspot area is far from being in agreement with a standard Gaussian (as an example, see the gaussian distribution highlighted at the bottom from Fig. \ref{fig1}) since the values of $q(A)$ differ significantly from unity in all situations (see also Table \ref{table2}).

According to Figure \ref{fig3} (top panel) the index $q_{N}$ is dominant over $q_{S}$. The same behavior can be verified when referring to the $\beta_{q}$-index as shown in the lower panel of the same figure. In particular, \cite{chowdhury} points out that the asymmetry of cycle 23 and the asymmetry phase of cycle 24 is predominantly dominated by the northern hemisphere. \cite{vizoso1990} mention that during the period from 1957 to 1970 the northern hemisphere is clearly dominant, but after 1970 this behavior is reversed. Our work extends this analysis by comparing the periods of rise and fall of the hemispheres considering the last 13 solar cycles. It is worth noting that the curves in Figure \ref{fig3} denote a possible phase transition that may be associated with the process of inversion of the solar magnetic dipole that occurs during the passage of each cycle. 

The curves highlighted in Fig. \ref{fig3} were obtained from a semi-sinusoidal fit also using the L-M algorithm. This adjustment was then used to obtain a harmonic best fit with the cyclic series of indexes $q$ and $\beta$ as follows:
\begin{equation}
\label{ss}
y(t)=a\sin\left[\frac{\pi(x-x_{a})}{P_{a}}\right]+b\cos\left[\frac{\pi(x-x_{b})}{P_{b}}\right]+c_{0},
\end{equation}
where $a$ and $b$ are amplitudes, $P_{a}$ and $P_{b}$ are periods, $x_{a}$ and $x_{b}$ are phase shifts and $c_{0}$ is the background level. The best-fitting parameters for the cycle–($q,\beta$) relationship using equation \ref{ss} are depicted in Table \ref{table1}.

    \begin{table}
     \caption{\footnotesize{Best parameter values of our semi-sinusoidal model using equation \ref{ss}. The values of $R^{2}$ also are shown.}}\label{table1}
      \begin{tabular}{ c c c c c }
        \\
        \hline	Parameters & $q_{N}$ & $q_{S}$ & $\beta_{N}$ & $\beta_{S}$ \\
        \hline
        %\multicolumn{2}{c}{<>}
        a	&	0.07$\pm$0.05	& -1.46$\pm$0.08 & 2.83$\pm$0.55 & 1.72$\pm$0.52 		\\
        b	&	0.12$\pm$0.09	&	1.59$\pm$0.10 & 2.04$\pm$0.79 & 2.32$\pm$0.57		\\
        $x_{a}$	&	-4.80$\pm$0.6	& 9.31$\pm$0.03 & -2.81$\pm$0.20 & 14.50$\pm$0.42		\\
        $x_{b}$	&	-5.39$\pm$0.4	&	0.69$\pm$0.03 & 3.14$\pm$0.90 & 6.80$\pm$0.45		\\
        $P_{a}$	&	2.85$\pm$0.74	& 2.53$\pm$0.41 & 2.72$\pm$0.3 & 1.58$\pm$0.11		\\
        $P_{b}$	&	2.06$\pm$0.24	&	2.49$\pm$0.50 & 1.99$\pm$0.14 & 2.24$\pm$0.07		\\
        $c_{0}$	&	2.17$\pm$0.02	&	2.11$\pm$0.02 & 8.71$\pm$0.38 & 7.39$\pm$0.39		\\
        $R^{2}$	&	0.82	&	0.76 & 0.75 & 0.86	\\
        \hline
      \end{tabular}
    \end{table}

We can explain the semi-sinusoidal variability of the $q$-index based on the area occupied by sunspots. In this perspective, the higher values denote a wide spectrum of variety about the size of the area, causing an extension in the tail of the distribution and, therefore, raising the value of the $q$-index. On the other hand, a smaller diversity in the spots' size reduces the tail's width, generating the lowest values of this index. Consequently, the periodic cycle is a manifestation of the complex dynamics that govern the life of the spots and the drop in photometric magnitude (directly proportional to the occupied area) caused by them. This result is corroborated by the parameters $\beta$ which follows the same trend as the index $q$. It is important to emphasize that during the 13 cycles, the index does not approach unity ($q=1$) in both hemispheres.

In both hemispheres, we find that the periods $P_{a}$ and $P_{b}$ are approximately twice the duration of the sunspot cycle (see also Table \ref{table1}). These periods are very close to the long-term magnetic field variation known as the 22-year Hale cycle. This behavior can be verified for both the $q$ and $\beta$ indexes \citep{KOTOV2015979} (see Fig. \ref{fig3}). It is noteworthy that this cycle cannot be explained by the canonical dynamo theory and its origin is associated with cosmic ray flux \citep{thomas}. In this way, the physical implications for the cyclical behavior of the entropic index $q$ can be linked to modulation by the heliospheric magnetic field and, consequently, it is also associated with the polarity-dependent effect of the solar magnetic field \citep{KOTOV2015979,thomas}. This issue deserves special attention and we will dedicate efforts in a forthcoming communication.

On this wise, it is possible to highlight the cyclical trend of each index $q$ and, therefore, the asymmetric inversion in a given cycle. \cite{vizoso1990} mention that the asymmetry inversion is found when the Sun's magnetic dipole inverts. This implication reinforces the idea that the $q$ index has a close correlation with the mechanisms that control the long-term variation of the solar dynamo. In other words, the preference of one hemisphere over the other suggests that the process of creating spots in the solar photosphere is notoriously non-ergodic following the rules of out-of-equilibrium statistical mechanics. 

As reported by \cite{chowdhury}, on small timescales of the order of a few months and a year or two, that is, for periodic fluctuations controlled by differential rotation and the meridional circulation, solar activity is generated independently in the two hemispheres. However, our results indicate that for long-term periodic fluctuations, such as the 11-year cycle, the conclusion of \cite{chowdhury} is not valid, i.e., the hemispheres appear to work together and not independently. A result in favor of this statement is given by the behavior of the index $q(A)$. According to Figure \ref{fig4}, the asymmetry also varies approximately cyclically, which may be a strong indication that the conclusion of \cite{chowdhury} is valid only for regimes that go from the period of solar rotation to the seasonal cycle of $\sim$ 1 year. 

Also according to Figure \ref{fig4}, the asymmetry indexes $A$ and $q(A)$ are more discrepant between cycles 12 to 14 and 22 to 24 because the preference between cycles is reversed. In the other cycles, the correlation trend follows, but with different amplitudes, mainly in cycles 19 and 20. In general terms, the difference can be explained by the fact that the asymmetry values $q(A)$ take into account the tail of the distribution, while the index $A$ takes only the average values of the sunspot area. This means that a complete view of the canonical index $A$ would depend on a more careful evaluation of the behavior of the standard deviation of the time series in each cycle. This requirement can be neglected in the case of $q(A)$ because it already provides information about how far the distribution is from the gaussian one. In this way, the index $q(A)$ brings us a new approach to the solar hemispheric asymmetry, indicating that the cyclicity does not depend only on the maximum and minimum values, but, above all, on the deviation of the Gaussianity that measures the presence of temporal autocorrelations (or memory effects) in the dynamics of the solar cycle.

In summary, our results reveal that a non-extensive interpretation for this context is necessary. Overall, the strength of the present study lies in the fact that the behavior of the asymmetric $q$ index places the northern hemisphere as the dominant one in most solar cycles.

\section{Concluding remarks}
Our work aimed to investigate how the nonextensive formalism can explain the N-S hemispheric asymmetry, especially why the asymmetry phase is mostly dominated by the northern hemisphere. To this end, we used a sample of 13 solar cycles divided by hemisphere extracted from the MSFC survey. 

The main conclusions drawn from this study are presented below, in the form of topics:

\begin{itemize}
\item The main conclusion obtained from the present dissertation refers to the entropic parameter $q$ as an efficient measure to diagnose long-term variations of the solar dynamo;
\end{itemize}
\begin{itemize}
\item The $q$ and $\beta_{q}$ parameters vary similarly. Both indicate that the northern hemisphere is dominant over the southern hemisphere;
\end{itemize}
\begin{itemize}
\item As indicated by Fig. \ref{fig3}, the cyclic variation of $q$-asymmetry may be associated with the process of inversion of the solar magnetic dipole that occurs during the 22-year Hale cycle;
\end{itemize}
\begin{itemize}
\item The preference of one hemisphere over the other shows that the process of creating spots in the solar photosphere is non-ergodic;
\end{itemize}
\begin{itemize}
\item Our results indicate that the conclusion mentioned by \cite{chowdhury} for short-term periodic variations is not valid for long-term ones;
\end{itemize}

\begin{itemize}
\item All the results found in the present study correspond to fluctuations between the nearest neighbors. For other scales, for example, of the order of the solar rotation period (23$-$34 days) or the lifetime of the flares (156 days), the behavior of the $q$ index for the present problem may be different. However, our results confirm that at the multifractal noise level ($\tau=1$ day) there are strongly non-extensive processes. A deeper analysis at different scales is necessary to verify if the behavior of $q$ is a function of the adopted scale and, even more, if there is a scale in which the regime drops to gaussianity as predicted by the Central Limit Theorem.
\end{itemize}
\begin{itemize}
\item Lastly, our study opens a new way to investigate the Sun from the perspective of out-of-equilibrium statistical models.
\end{itemize}

In addition, it is worth mentioning that the entropic index $q$ can also be used to investigate other possible sources for the hemispheric asymmetry. In this sense, a deeper analysis must be done on solar indicators such as magnetic field and Total Solar Irradiance (TSI) from Virgo/SoHO data and X-ray flux. This issue will be addressed in a forthcoming communication.

%%%%%%%%%%%%%%%%%%%%%%%%%%%%%%%%%%%%%%%%%%%%%%%%%%%%%%%%%%%%%%%%%%%%%%%%%%%
%% Appendix
%
% \appendix

%%%%%%%%%%%%%%%%%%%%%%%%%%%%%%%%%%%%%%%%%%%%%%%%%%%%%%%%%%%%%%%%%%%%%%%%%%%
%% Acknowledgements
%
 \begin{acks}
    DBdeF acknowledges financial support 
    from the Brazilian agency\\ CNPq-PQ2 (Grant No. 305566/2021-0). Research activities of STELLAR TEAM of Federal University of Cear\'a are supported by continuous grants from the Brazilian agency CNPq. This paper includes data collected by the NASA’s Marshall Space Flight Center (MSFC). Supporting for this database is provided by the NASA Science Mission directorate. Funding for this database terminated in FY2005 (last update 2017/03/23). All data presented in this paper were obtained from the Royal Observatory, Greenwich - USAF/NOAA Sunspot Data. Data of our analyses presented in this paper will be shared on reasonable request to the corresponding author.

 \end{acks}

%% Available additional data environments:
%% required: authorcontribution, fundinginformation, dataavailability
%% optional: materialsavailability, codeavailability
% \begin{authorcontribution}
%
% \end{authorcontribution}
%
% \begin{fundinginformation}
%
% \end{fundinginformation}
%
% \begin{dataavailability}
%
% \end{dataavailability}
%
% \begin{ethics}
% \begin{conflict}
%
% \end{conflict}
% \end{ethics}

%%% %%%%%%%%%%%%%%%%%%%%%%%%%%%%%%%%%%%%%%%%%%%%%%%%%%%%%%%%%%%
%% Bibliography
%
% Using BibTeX
%
\bibliographystyle{spr-mp-sola}
\bibliography{qSolarAsymmetry2022}

%
% Without BibTeX
% \begin{thebibliography}{}
% \bibitem[\protect\citeauthoryear{Author}{Year}]{key}
%   <bibliographical entry>
%
% \bibitem[\protect\citeauthoryear{}{}]{}
%
%\begin{thebibliography}{99}
%\bibitem[\protect\citeauthoryear{author}{year}]{label} 
%\end{thebibliography}

\end{article}
\end{document}